\documentclass[11pt]{article}

\usepackage{tikz,enumerate}
\usepackage[framemethod=TikZ]{mdframed}
\usetikzlibrary{arrows}
\newdimen\arrowsize
\pgfarrowsdeclare{arcsq}{arcsq}
{
  \arrowsize=0.2pt
  \advance\arrowsize by .5\pgflinewidth
  \pgfarrowsleftextend{-4\arrowsize-.5\pgflinewidth}
  \pgfarrowsrightextend{.5\pgflinewidth}
}
{
  \arrowsize=1.5pt
  \advance\arrowsize by .5\pgflinewidth
  \pgfsetdash{}{0pt} 
  \pgfsetroundjoin   
  \pgfsetroundcap    
  \pgfpathmoveto{\pgfpoint{0\arrowsize}{0\arrowsize}}
  \pgfpatharc{-90}{-140}{4\arrowsize}
  \pgfusepathqstroke
  \pgfpathmoveto{\pgfpointorigin}
  \pgfpatharc{90}{140}{4\arrowsize}
  \pgfusepathqstroke
}

\usepackage{pgf,pgfarrows,pgfnodes,pgfautomata,pgfheaps,pgfshade}

\usepackage{fullpage}
\usepackage{graphicx}
\usepackage{latexsym,amsmath,amssymb,amsthm,epic,eepic,multirow}
\usepackage{natbib}

\usepackage{multirow}
\usepackage{subfigure}
\usepackage{natbib}
\usepackage{algorithm}
\usepackage{algorithmic}
\usepackage{url}

\usepackage{dsfont}

\newcommand{\EE}{\mathbb{E}}
\newcommand{\R}{\mathbb{R}}

\newcommand{\eps}{\varepsilon}

\newcommand{\Cov}{\mbox{Cov}}

\newcommand{\Var}{\mbox{Var}}

\newcommand{\argmin}{\mathrm{argmin}}

\newtheorem{fact}{Fact}

\usepackage{color}

\begin{document}

\title{Deconfounding and Causal Regularization\\
  for Stability and External Validity} 

\author{Peter B\"{u}hlmann and Domagoj \'Cevid\\
Seminar for Statistics, ETH Z\"urich}
 
\date{\today}

\maketitle

\begin{abstract}
We review some recent work on removing hidden confounding and causal
regularization from a unified viewpoint. We describe how simple and
user-friendly techniques improve stability, replicability and
distributional robustness in heterogeneous data. In this sense, we provide
additional thoughts to the issue on concept drift, raised by
\citet{efron2020}, when the data generating distribution is changing.  
\end{abstract}

\section{Introduction}

Brad Efron, in his lecture at the occasion of receiving the International
Prize in Statistics, brought up some fascinating thoughts on
``prediction, estimation and attribution'', with particular attention to the
new ``wide data era'' which has entered statistics and data science more
generally \citep{efron2019,efron2020}.
Looking back almost 20 years ago, there has been a huge
development in statistics since Leo Breiman's article ``Statistical
Modeling: The Two Cultures'' \citep{breiman2001statistical}. Even more
broadly, data science has become an emerging new field and
profession. It deals with information extraction from data, often in
close proximity with other sciences. Its historical roots are in statistics, and
statistical ``critical'' thinking plays an ever important role in inference
from data to models and prediction. 
There are many interesting facets of this broad topic, see for example
David Donoho's ``50 years of Data Science'' \citep{donoho201750} or Bin
Yu's ``Veridical Data Science'' \citep{yu2020veridical}.

\citet{efron2019,efron2020} has formulated intriguing ideas on
``prediction, estimation and attribution''. 
We are presenting here a few additional considerations on the topic, as
outlined in the following Sections \ref{subsec.stab} and
\ref{subsec.extval}. 

\subsection{Stability of predictions and causal thinking in presence of perturbations: Efron and Cox in response to \citet{breiman2001statistical}}\label{subsec.stab}  
\citet{breiman2001statistical} argued strongly in favor of
prediction and the corresponding feature importance measures. However,
prediction in reality is often more subtle than the usual textbook
definition where one assumes the same data generating mechanism for the
training and the new test set data.
 
The illustration by \citet{efron2020} of concept drift where the
data-generating distribution changes between training and test set, 
or also
his question ``Were the test sets really a good test?'' \citep{efron2019}, 
nicely emphasizes that prediction can be ``highly
context-dependent and fragile'': he illustrates with a certain dataset that
training on the first part of the observations 
and using the last ones as the test set gives a widely different answer for the
error rate than the
average of taking many random divisions into training- and
test-data. Apparently, the last observations in the dataset seem to have a
rather different data generating distribution than the first ones from the
training phase. 

Similarly, \citet{cox2001statistical} wrote in a response to Breiman's
article wrote: 
``... Key issues are then the stability of the predictor as practical prediction
proceeds, the need from time to time for recalibration and so on. However,
much prediction is not like this. Often the prediction is under quite
different conditions from the data; what is the likely progress of
the incidence of the epidemic of v-CJD in the United Kingdom, what would be
the effect on annual incidence of cancer in the United States of reducing
by 10\% the medical use of X-rays, etc.? That is, it may be desired to
predict the consequences of something only indirectly addressed by the data
available for analysis. As we move toward such more ambitious tasks,
prediction, always hazardous, without some understanding of underlying
process and linking with other sources of information, becomes more and
more tentative. Formulation of the goals of analysis solely in terms of
direct prediction over the data set seems then increasingly unhelpful.''

Whereas \citet{efron2001statistical} wrote in return to Breiman's article:
``Estimation and testing are a form of prediction: ``In our sample of 20
patients drug A outperformed drug B; would this still be true if we went on
to test all possible patients?'' ...(Peter Gregory) undertook his study for 
prediction purposes, but also to better understand the medical basis of
hepatitis. Most statistical surveys have the identification of causal
factors as their ultimate goal.''

In this paper, we build on the fact that stability of prediction and
causality are naturally connected. As a result, new methods and
algorithms emerge 
which are easy to use and fairly ``automatic''. They will not replace
careful statistical thinking, for example in the way
\citet{cox2001statistical} describes it above. But they often act, in quite
a few scenarios, more intelligently 
than plain vanilla ``black box'' prediction algorithms: perhaps, such and
many other new algorithms close to some extent the gap between ``the
two cultures'' from \citet{breiman2001statistical}. This is somewhat in
line with Brad Efron's statements in his International Prize in Statistics
lecture \citep{efron2019}, namely ``Two Trends: Making prediction algorithms
better for scientific use'' and ``Making traditional estimation/attribution
methods better for large-scale $(n;p)$ problems''. 
 
\subsection{External validity, distributional replicability, robustness and connections to causality}\label{subsec.extval}  

One major problem with many modern algorithms and methods
is their vulnerability to distributional changes in new data. Would we see
a good amount of replication 
in a new study, or in a new environment? Can we do accurate
prediction and estimation in changing scenarios? These questions tie in to
some of the
points raised by \citet{efron2020} and mentioned above, namely about
concept drift 
(``Were the test sets really a good test?'' \citep{efron2019}), or to the
comments by \citet{cox2001statistical} that 
``direct prediction over the data set seems then increasingly unhelpful''. 
They both refer to external validity and generalization
beyond the observed data. The latter is well understood if the future external
data has the same generating distribution as the observed training data,
but if not, external validity relates to distributional robustness
\citep{sinha2017,gao2017,meinshausen2018}, transfer learning
\citep{pratt1993,pan2010survey} and
causality \citep{diddaw10,jonipb16,bareinboim2016causal,rojas2018,rothenhetal18,pb20,daw20}.  
%

%
%
%
%

\subsection{The current work}

We review some of our more recent contributions on deconfounding, distributional
robustness and replicability, and causality
\citep{rothenhetal18,pb20,cevid2018spectral,guoetal2020}. A unified
treatment might enable us to clarify the connections more clearly. We aim for
simplicity, 
demonstrating that at least some of the ideas and methods are simple and
easy to use, yet they seem to be effective in achieving some form of
distributional robustness. The latter term is rather different from the
more standard formulation and procedures in robust statistics
\citep{huber1964,hampeletal86}, where outliers occur in the training data
and unlike test set distributional changes examined here. 

\begin{figure}[!htb]
\begin{center}
\begin{tikzpicture}[xscale=2.5, yscale=1.2, line width=0.5pt, minimum size=0.58cm, inner sep=0.3mm, shorten >=1pt, shorten <=1pt]
    \normalsize
    \draw (1,0) node(x) [circle, draw] {$X$};
    \draw (3,0) node(y) [circle, draw] {$Y$};
    \draw (0.5,2) node(a) [circle, draw] {$A$};
    \draw (2,2) node(h) [circle, draw] {$H$};
    \draw[-arcsq] (x) -- (y);
    \draw[-arcsq] (y) -- (x);
    \draw[-arcsq] (a) -- (x);
    \draw[-arcsq] (h) -- (x);
    \draw[-arcsq] (x) -- (h);
    \draw[-arcsq] (h) -- (y);
    \draw[-arcsq] (y) -- (h);
    \draw[-arcsq] (a) -- (y);
    \draw[-arcsq] (a) -- (h);
    \draw (2,0.2) node(e) {$\beta^0$ };
    \draw(0.3,2.5) node(f) {perturbations};
    \draw(2,2.5) node(k) {hidden confounders};
  \end{tikzpicture}
  \caption{The generic problem. The goal is inferring the regression
    parameter $\beta^0$, describing the
    relation between (the causal components of) $X$ and $Y$. Additionally,
    there are hidden confounding variables $H$ and 
    perturbations generated by observed external (exogenous) variables $A$ are
    present. The cases without and with $A$ are discussed in Sections
    \ref{sec.deconfounding} and \ref{sec.anchor}, respectively. 
    The directionality among different variables may be
    unknown. The variables $X, H, A$ can be multivariate but for
    simplicity, $Y$ is univariate. The graph corresponds to the structure of a
    structural equation model: the arrows are bi-directed, saying that the
    directions between some of the components can go either way and feedback
    loops are allowed as well.}\label{fig1}
\end{center}
\end{figure}
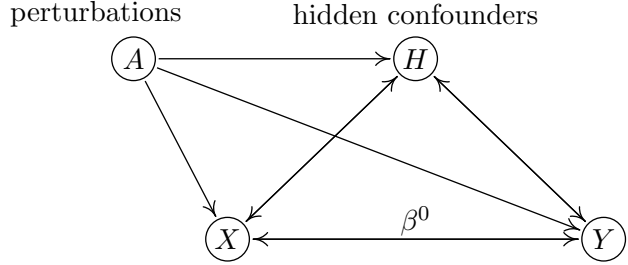

The generic problem we are considering is loosely illustrated in Figure
\ref{fig1}. We are interested in inferring the unconfounded regression
parameter  $\beta^0$ and in stable
prediction of $Y$ from 
$X$. We do not observe all the relevant variables and are thus confronted
with hidden confounding. This scenario is discussed in Section
\ref{sec.deconfounding}.  
Additionally, we may observe data under various
perturbations which are generated by external (exogenous) variables
$A$, as discussed in Section \ref{sec.anchor}.
The graph in Figure \ref{fig1} corresponds to a structural equation
model \citep{bollen89,pearl00}, introduced in equation
\eqref{SEMlin} or \eqref{mod.anchor} more formally. Of particular interest is
the univariate response $Y_i$ and its linear regression function of some of the components of a $(1 \times p)$-dimensional
covariate $X_i$, where $i$ denotes the $i$th observation:
\begin{eqnarray*}
  Y_i \leftarrow X_i \beta^0 + g(H_i,A_i) + \eps_{Y,i},
\end{eqnarray*}
where $\eps_{Y,i}$ is a noise or innovation term being independent of all the
variables arising ``earlier'' or ``up-stream'' of $Y_i$; the exogenous
variables $A_i$ are non-existent in our discussion in Section \ref{sec.deconfounding}.  
The symbol ``$\leftarrow$'' is algebraically an equality sign. 
The variables
corresponding to the support of $\beta^0$ are the \emph{causal
  $X$-variables for $Y$}, since they are the only components of $X_i$ which
directly enter the structural equation for $Y_i$.  Thus, $\beta_j^0 \neq 0$
if and only if the $j$th component of $X$ corresponds to a causal
$X$-variable. More precise definitions of the model versions are given 
later. 

\paragraph{A connection to causality.} 
There is a fruitful link to causality. In a nutshell, one can represent the
causal parameter $\beta^0$ (or also the function $f^0$) as the minimizer of
a worst case risk such as
\begin{eqnarray*}
  \beta^0 = \mbox{argmin}_{\beta} \max_{P \in {\cal P}} \EE_P[(Y_i - X_i
  \beta)^2],
  \end{eqnarray*}
for particular classes of distributions ${\cal P}$ for $(X,Y)$. Such a
class can be 
thought as containing various perturbations of the original data generating
distribution and hence, there is an intrinsic connection between
causality and distributional robustness \citep{diddaw10,jonipb16, rojas2018,
  rothenhetal18, pb20, daw20}. In this paper, we will not elaborate much on
the causal interpretation: however, the operational procedures which have causal
interpretability can be ``simply'' used to increase robustness and the
degree of external validity. 

\paragraph{Notation.} We use the standard notation in regression or
classification and denote by $X$ and $Y$ the
observed $n \times p$ design matrix of covariates and the $n \times 1$
response vector of the data, respectively; $n$ is the sample size and $p$
the dimensionality of the covariates. The $i$th instance is denoted by
$X_i$ and $Y_i$, respectively, with $X_i$ being a $1 \times p$ vector. 

\section{Deconfounding: in presence of dense confounding}\label{sec.deconfounding}

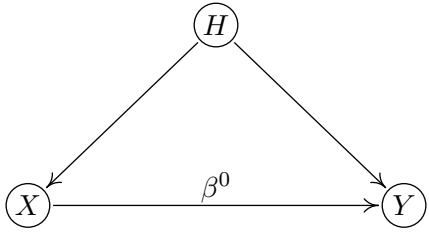
\begin{figure}[!htb]
\begin{center}
\begin{tikzpicture}[xscale=2.5, yscale=1.2, line width=0.5pt, minimum size=0.58cm, inner sep=0.3mm, shorten >=1pt, shorten <=1pt]
    \normalsize
    \draw (1,0) node(x) [circle, draw] {$X$};
    \draw (3,0) node(y) [circle, draw] {$Y$};
    \draw (2,2) node(h) [circle, draw] {$H$};
    \draw[-arcsq] (x) -- (y);
    \draw[-arcsq] (h) -- (x);
    \draw[-arcsq] (h) -- (y);
    \draw (2,0.2) node(e) {$\beta^0$};
  \end{tikzpicture}
  \caption{Structure of the linear model with unobserved confounding
    variables $H$ as in \eqref{SEMlin}.}\label{fig2}
\end{center}
\end{figure}

We consider the well-known problem of unobserved hidden confounding in a
regression context. This is a special case of Figure \ref{fig1}, where the
directions are known and without perturbations from external (exogenous)
variables, see Figure \ref{fig2}. 

There are several ways to explain it: we do so by using structural equation
models (SEMs), see for example \citet{bollen89} or \citet{pearl00}. We
observe a univariate response 
variable $Y_i$, a $(p\times 1)$-dimensional covariate $X_i$ and an
unobserved $(q \times 1)$-dimensional 
hidden confounding variable $H_i$.
In the linear case, the model is set up as follows:
\begin{eqnarray}\label{SEMlin}
  & &Y_i \leftarrow X_i \beta^0 + H_i \delta + \eps_{Y,i},\nonumber\\
  & &X_i \leftarrow H_i \gamma + \eps_{X,i},\nonumber\\
  & &\eps_{X,i},\eps_{Y,i}, H_i\ \mbox{jointly independent},
\end{eqnarray}
where
$\beta, \delta$ are column vectors and $\gamma$ a $q \times p$ matrix.
We typically make an i.i.d. assumption across the indices
$i=1,\ldots ,n$.  
The symbol ``$\leftarrow$'' is algebraically an equals sign and it means in
addition that the factorization of the joint distribution of (all
components of) $Y_i,X_i,H_i$, namely $p(y,x,h) = 
p(y|x,h) p(x|h) p(h)$ with conditional distributions (densities), is
precisely described by the equations, e.g., $p(x|h) = p_{\eps_X}(x -
h\gamma)$.   
Figure \ref{fig2} shows the corresponding graphical structure of the model. Of
particular interest is the equation for the response variable $Y$: the goal
is to infer the parameter $\beta^0$ from data. In the causality literature,
$\beta^0$ is called the causal parameter of $X_i$ on $Y_i$; but even without
using the word causality, we can view it as the ``internal systems parameter''. 

The parameter $\beta^0$ is not the regression parameter of $Y$ versus
$X$. In fact, due to confounding by the unobserved confounding variables
$H_i$, we have:   
\begin{eqnarray*}
  & &\mbox{argmin}_{\beta}\ \EE[(Y_i - X_i \beta)^2] = \beta^0 + b,\\
  & &b = \Cov(X_i)^{-1} \Cov(X_i,H_i) \delta. 
\end{eqnarray*}
We can thus represent the model for the $Y$ equation in \eqref{SEMlin} as a
standard linear model
\begin{eqnarray}\label{linmod}
  Y_i = X_i (\beta^0 + b) + \eps_i,\ \eps_i = (H_i \delta - X_i b) +
  \eps_{Y;i},
\end{eqnarray}
where $\eps_{i}$ is uncorrelated with $X_i$, due to the property of the
$L_2$ projection. A remarkable fact is that the bias $b$ becomes small in
case of high-dimensionality and ``dense'' confounding as explained next. 

\medskip\noindent
\emph{A simple example of dense confounding.}\\
Consider the case for equation \eqref{SEMlin} where $H$ is
1-dimensional with $\Var(H_i) = 1$ and $\Cov(\eps_{X;i}) = \xi^2 I$: $b =
(\gamma^T \gamma + \xi^2 I)^{-1} \gamma^T \delta$,
and for $\xi^2 \ll \|\gamma\|_2$ (in the context of dense confounding,
$\|\gamma\|_2 \asymp \sqrt{p}$, see below) we obtain that 
\begin{eqnarray*}
  \|b\|_2 \approx |\delta|/\|\gamma\|_2.
\end{eqnarray*}
Hence, if, say, all components of $\gamma$ are of order one, that is, every
component of $X_i$ is affected by $H_i$ with size of order one,  which is some
kind of dense confounding, we have that $\|b\|_2 =
\mathcal{O}(|\delta|/\sqrt{p})$. Therefore, this is a 
blessing of dimensionality when $p$ is large. 

\medskip
One can see from the above example that the bias term of the population
least squares principle becomes small in the case of high-dimensionality
and dense 
confounding. However, with estimation based on finite sample size, several
issues become more delicate and we propose to modify penalized least
squares methods, as discussed next.  
%
%

\subsection{Deconfounding with spectral transformations}\label{subsec.spectrans}

For estimating $\beta^0$ in \eqref{SEMlin} we use a simple
pre-processing technique which has some mathematical guarantees under an
additional assumption of dense confounding.

\paragraph{Principal component adjustment.}
As a motivation, we consider first a commonly used approach to guard against
hidden confounding as in \eqref{SEMlin}. We extract the first few
principal components of $X$, denoted by 
$W^{(1)},\ldots W^{({\hat{q}})}$, ideally with $\hat{q}$ equal to or
slightly larger than
$q$. The $(n \times \hat{q})$ principal components $W = (W^{(1)},\ldots
,W^{(\hat{q})})$ 
serve as a proxy for the unobserved $(n \times q)$ $H$: the approximation
is reasonable if 
the orthogonal projection $\Pi_W = W (W^T 
W)^{-1} W^T$ is similar to $\Pi_H = H (H^TH)^{-1} H^T$, that is, if $X$ has
approximately a low rank structure. One then adjusts for
the principal components in $W$ and builds partial residuals:
\begin{eqnarray}\label{pcaadj}
  \tilde{X} = (I - \Pi_W) X,\ \ \tilde{Y} = (I - \Pi_W) Y
  \end{eqnarray}
and proceeds with (regularized) least squares regression of $\tilde{Y}$
versus $\tilde{X}$ to estimate the parameter $\beta^0$ in \eqref{SEMlin}. We
can interpret this procedure in terms of singular values of $X$. Let
\begin{eqnarray*}
  X = UDV^T
\end{eqnarray*}
be the singular value decomposition (SVD) of $X$. The singular values $D =
\mbox{diag}(d_1,\ldots ,d_m)$, with $m = \min(n,p)$ are ordered as $d_1 \ge
d_2 \ge \ldots \ge d_m$. Consider a truncation of the singular
values to
\begin{eqnarray}\label{pcashrink}
  \tilde{d}_{\mathrm{PCA},i} = 0\ (i=1,\ldots ,\hat{q}),\
  \tilde{d}_{\mathrm{PCA},i} = d_i\ (i=\hat{q}+1,\ldots ,m). 
\end{eqnarray}
The PCA-adjusted matrix $\tilde{X}$ in \eqref{pcaadj} can then be written
as $\tilde{X} = U \tilde{D} V^T$, where $\tilde{D} =
\mbox{diag}(\tilde{d}_1,\ldots, \tilde{d}_m)$ (note that $W^{(1)}\ldots ,
W^{(\hat{q})}$ are, when standardized to unit length, the first $\hat{q}$
column vectors of $U$). Alternatively, we can
represent $\tilde{X}$, and also $\tilde{Y}$ in \eqref{pcaadj} as a linear
spectral transformation of the original quantities:
\begin{eqnarray}\label{pcaspec}
  & &\tilde{X} = FX,\ \ \tilde{Y} = FY,\nonumber\\
  & &F = U \mbox{diag}(\tilde{d}_{\mathrm{PCA},1}/d_1,\ldots
      ,\tilde{d}_{\mathrm{PCA},m}/d_m) U^T, 
\end{eqnarray}
and of course, we then have that $F = I - \Pi_W$. 

\subsubsection{The Trim transform, and relations to Lava}

\begin{figure}[!htb]
  \begin{center}
    \includegraphics[scale=0.1]{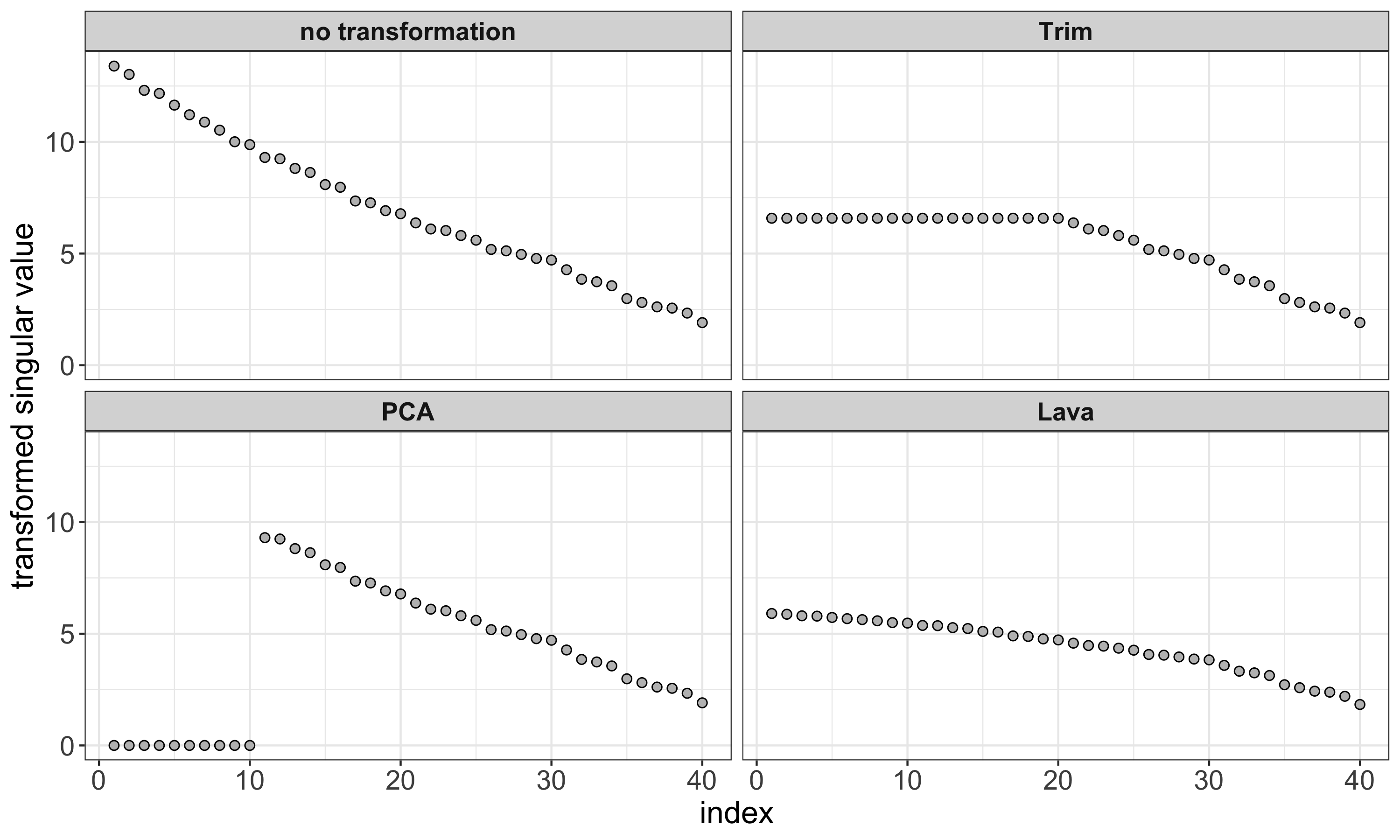}
    \caption{Singular values of spectral transformed $\tilde{X}$. From
      top left to bottom right: no transformation with original $X$ matrix,
    Trim-transform, PCA adjustment with 10 principal components,
    Lava. The figure is essentially taken from
    \citet{cevid2018spectral}.}\label{fig.transf}  
  \end{center}
  \end{figure}

One can think of other data transformations than the one in \eqref{pcaadj} or
\eqref{pcaspec}.
In fact, one may ask the question why the largest singular values in
\eqref{pcashrink} are shrunken to zero, making them the smallest singular
values in the transformed $\tilde{X}$, see Figure \ref{fig.transf}.
It
might be advantageous
to keep the ordering of singular values in the transformed design matrix
while still shrinking the large ones. Two particular choices are as
follows. The Trim 
transform \citep{cevid2018spectral} uses 
\begin{eqnarray*}
  \tilde{d}_{\mathrm{Trim},i} = \min(d_i,\tau),\ i= 1,\ldots ,m,
\end{eqnarray*}
where $\tau$ is some threshold. A generic and often very good value is
$\tau = d_{\lfloor m/2 \rfloor}$, the median of the singular values, see
Figure \ref{fig.transf}.
The transformed variables are then of exactly the same form as in
\eqref{pcaspec}, namely pre-multiplying with the linear transformation $F$,
involving now $\tilde{d}_{\mathrm{Trim},i}$ being different than in
\eqref{pcashrink}. Once we have the transformed data $\tilde{X}$ and
$\tilde{Y}$, one can use ``any'' reasonable technique for (high-dimensional)
linear regression, say the Lasso \citep{tibs96}
\begin{eqnarray}\label{TrimLasso}
  \hat{\beta}_{\mathrm{TrimLasso}} = \mbox{argmin}_{\beta} \|\tilde{Y} - \tilde{X} \beta\|_2^2/n
  + \lambda \|\beta\|_1,
\end{eqnarray}
where $\lambda > 0$ is the regularization parameter. The special case of
ordinary least squares with $\lambda = 0$ is briefly mentioned in footnote
1. 
Other sparse estimators than the Lasso are possible as well, such as
forward \citep{efroymson1960multiple} or stagewise selection
\citep{efron04lars}, Elastic Net \citep{zouhastie05}, or regularization
with the SCAD
\citep{fanli01} or MCP \citep{zhang2010} penalty. 
We will describe in Fact \ref{th1} that the Lasso with the Trim
transform $\hat{\beta}_{\mathrm{TrimLasso}}$ 
estimates the parameter $\beta^0$ in \eqref{SEMlin}, assuming dense
confounding. We note in particular, that the construction of the
estimator is \emph{very simple and easy to use,} requiring no further
specialized software. 

Another choice of a spectral transformation as in \eqref{pcaspec} is
implicit in the Lava \citep{chernozhukov2017lava} estimator with
\begin{eqnarray*}
  \tilde{d}_{\mathrm{Lava},i} = \sqrt{\frac{n \lambda_2 d_i^2}{n
  \lambda_2 + d_i^2}},
\end{eqnarray*}
where $\lambda_2 > 0$ is a tuning parameter. It is
argued in \citet{cevid2018spectral} that the choice
\begin{eqnarray}\label{lambda2lava}
  \lambda_2 = d^2_{\lfloor m/2
  \rfloor}/n
\end{eqnarray}
is similar to the Trim transform with $\tau = d_{\lfloor m/2
  \rfloor}$, see also Figure \ref{fig.transf}. We just point out that the
Lava spectral 
transformation has an interesting representation in terms of estimating
$\beta^0$ in \eqref{SEMlin}. It holds algebraically that for
\begin{eqnarray*}
(\hat{\beta},\hat{b})_{\mathrm{Lava}} = \mbox{argmin}_{\beta,b} \|Y - X(\beta + b)\|_2^2/n
  + \lambda_1 \|\beta\|_1 + \lambda_2 \|b\|_2^2,
\end{eqnarray*}
we can represent, as in \eqref{TrimLasso},
\begin{eqnarray*}
  \hat{\beta}_{\mathrm{Lava}} = \mbox{argmin}_{\beta} \left(\|\tilde{Y} -
  \tilde{X} \beta\|_2^2/n + \lambda \|\beta\|_1 \right),
\end{eqnarray*}
where $\tilde{X}$ and $\tilde{Y}$ are spectral transformed original
quantities as in \eqref{pcaspec} but with $\tilde{d}_{\mathrm{Lava}}$ above. 
In view of the representation in \eqref{linmod} with a sparse plus
dense parameter 
vector, the Lava estimator indeed estimates the sparse part $\beta^0$.

\subsection{Guarantees for the Lasso after the Trim transform}\label{subsec.Trimlasso}

Once we have the Trim-transformed data
$\tilde{X}$ and $\tilde{Y}$, we can use linear regression techniques for
estimating $\beta^0$ in \eqref{SEMlin}. We note that for least squares
estimation with $\mbox{rank}(X) = p < n$, nothing will
happen.\footnote{In fact one could use arbitrary values for $\tilde{D}$ as
  long as they are strictly positive.}\label{foot1} But for higher
dimensions and penalized methods, things change.  


We consider the Lasso on the Trim-transformed data as in \eqref{TrimLasso}
for some regularization parameter $\lambda$. Standard software can be used, 
for example \texttt{glmnet} in \texttt{R} \citep{friedetal09}. The choice
of the regularization 
parameter is 
perhaps a bit more delicate but we propose the usual e.g. 10-fold
cross-validation, see also below in Section
\ref{subsec.regularization}. This simple combination of 
deconfounding with the Trim transform in conjunction with the Lasso has
interesting theoretical guarantees under the following main assumptions:
\begin{description}
\item[(A1)] $\lambda_{\max}(\Cov(X_i,H_i)) \asymp \sqrt{p}$: the largest
  singular 
  value of the $(p \times q)$ covariance matrix of $(X_i,H_i)$ is of the
  order $\sqrt{p}$.     
\item[(A2)]$d_{\lfloor{n/2} \rfloor} = {\mathcal O}_P(\sqrt{p})$: the
  median value of the singular value of $X$ is 
  of the order $\sqrt{p}$, with high probability.
  \item[(A3)] The compatibility constant of $n^{-1} \tilde{X}^T \tilde{X}$
    is of the same order as the minimal eigenvalue
    $\lambda_{\min}(\Sigma)$ of $\Sigma = \Cov(X_i)$. 
  \end{description}
  \citet{cevid2018spectral} give a detailed discussion when these
  assumptions hold, see also \citet{guoetal2020}. In particular, (A1) is an
  assumption on dense 
confounding: for example, if $p/q \to \infty$ and the number of non-zero
columns of $\gamma$ is of the order $p$ (order $p$ components of $X$ are
affected by $H$) and each of the non-zero columns of $\gamma$ is
sampled i.i.d. from a sub-Gaussian vector, then (A1) holds with high
probability. This is an extension and along the lines of our simple example
above on dense confounding. Assumption (A2) holds with high
probability if the rows of $X$ are realizations of i.i.d.
random vectors (assuming sufficiently many finite moments). 
  \begin{fact}\label{th1}\citep{cevid2018spectral}
Consider the confounding model in \eqref{SEMlin} with $p \ge n$ and $\max_j
\Sigma_{jj} = \mathcal{O}(1)$, where $\Sigma = \Cov(X)$. Assume (A1)--(A3). Then, for some $\lambda
\asymp \sqrt{\log(p)/n}$ in \eqref{TrimLasso}, 
the usual rate of convergence as in the unconfounded high-dimensional
linear model holds, namely
\begin{eqnarray*}
  \|\hat{\beta}_{\mathrm{TrimLasso}} - \beta^0\|_1 = \mathcal{O}_P \left(\frac{\sigma
  s_0}{\lambda_{\min}(\Sigma)}\sqrt{\log(p)/n} \right),
\end{eqnarray*}
where $s_0 = |\mbox{supp}(\beta^0)|$ is the number of non-zero
components of $\beta^0$ and $\sigma^2 = \Var(H_i \delta + \eps_{Y;i})$.  
\end{fact}
The asymptotics is to be understood as the usual one in high-dimensional
statistics where both $p \ge n \to \infty$. 

\bigskip
Other methods such as forward selection \citep{efroymson1960multiple},
regularization with the SCAD
\citep{fanli01}, MCP \citep{zhang2010} or guaranteed
$\ell_0$ \citep{bertsimas2016best} based on 
the Trim-transformed data have not yet been theoretically established to
exhibit certain convergence rates. Fact \ref{th1} above serves as an
indication that algorithms and methods are expected to behave well when using
them on Trim-transformed data.

\subsection{Choosing the regularization
  parameter}\label{subsec.regularization} 

Choosing the regularization parameter for Lasso or other algorithms with
cross-validation is conceptually somewhat different 
than in the standard setting with no confounding.

For the sake of illustration, consider the Lasso
$\hat{\beta}_{\mathrm{TrimLasso}}(\lambda)$ on the Trim-transformed
data as in \eqref{TrimLasso}. When using cross-validation, aiming for best
prediction, the chosen $\lambda$ would be typically too small since the best
prediction would also try to capture the unwanted signal component $X b$ in
\eqref{linmod}. To partially correct for this issue,
cross-validation should 
be run on the deconfounded data $\tilde{X},\tilde{Y}$ and ignoring the
issue that the 
spectral transformation has used the full data; that is, we simply
spectral-transform the full data set first and then proceed as usual. This
strategy should make the additional signal $\tilde{X} b$ smaller and hence
cross-validation aiming for best prediction is expected to perform
reasonably fine. 

As an alternative to cross-validation, one can use Stability Selection
\citep{mebu10} on the original data. This
amounts to directly choosing an amount of 
regularization for selecting the relevant components of $\beta^0$, that is,
for variable selection. It does \emph{not} lead to an estimate for the
tuning parameter $\lambda$ in \eqref{TrimLasso}. Instead, Stability Selection is
linking a different stability-based regularization with the expected number
of false positives, 
assuming an exchangeability condition for i.i.d. generated data. However,
the methodology aiming for stability is also useful for heterogeneous data
where the underlying distribution has changed as discussed in next. 

\subsection{Robustification against hidden confounding and external validity}\label{subsec.robustify}

Perhaps the main value of the deconfounded Lasso procedure, i.e.,
Trim-transforming the data and using Lasso, is the degree of
robustification against hidden confounding. The assumptions (A1)--(A3) in
Section \ref{subsec.Trimlasso} might be partially unrealistic: but
\citet{cevid2018spectral} report 
empirically that ``there is not much to lose, but potentially a lot to be
gained''. This can be summarized as follows: (i) the procedure is extremely
simple requiring in addition only one SVD and (typically) three lines of
code; (2) the method is very effective in estimating the underlying
unconfounded regression parameter $\beta^0$ in scenarios of dense confounding
and a sparse $\beta^0$; (3) in case of no confounding, the
deconfounded Lasso is essentially as good as plain Lasso; (4) in between
the settings in (2) and (3), there is improvement with the deconfounded
Lasso over its plain version, yet it still does not entirely remove the bias
due to confounding. We refer also to Section \ref{subsec.fails}.   

The unconfounded parameter $\beta^0$ is the parameter
where other sources of unmeasured variation have been removed. This is very
relevant for improving replicability. Suppose that we estimate the
regression parameter on one (training) dataset and would like to have it
replicated on 
another (test) dataset. If the two datasets differ in their distribution,
the regression parameter is not replicable. However, the unconfounded
parameter $\beta^0$ is replicable under the following assumption:
\begin{eqnarray*}
\mbox{the training dataset}:& &\mbox{is generated from the model in}\
      \eqref{SEMlin},\\
 \mbox{the test dataset satisfies}:& &Y_i' \leftarrow X_i' \beta^0 + H_i'
                                       \delta' + \eps_{Y,i}',\nonumber\\ 
  & &X_i' \leftarrow H_i' \gamma' + \eps_{X,i}',\nonumber\\
  & &\eps_{X,i}',\eps_{Y,i}', H_i'\ \mbox{jointly independent},
\end{eqnarray*}
where the unconfounded parameter $\beta^0$ is the same but the other
parameters are allowed to change; the notation with the superscript $'$
denotes the quantities corresponding to the test dataset (but $\beta^0$ in
the test data is the same as in \eqref{SEMlin}). 
We will illustrate such a replicability phenomenon on real data below.

\subsubsection{An illustration on data from the GTEx consortium}\label{subsec.GTEx1}

The Genotype-Tissue Expression (GTEx) project is studying tissue-specific
gene expression and regulation in human
samples (\url{http://gtexportal.org}, \citet{lonsdale2013genotype}). Here,
we consider a  small aspect of the publicly available data.

For the specific skeletal muscle tissue, we have $14'713$ gene expression
measurements for $n = 491$ samples. In addition, there are 65 additional
covariates which are believed to be good proxies of confounding variables,
including genotyping 
principal components and so-called PEER factors. Thus, we have two data
sets: the raw data with covariates $X$ and response $Y$, and another with $X'$
and $Y'$, where we linearly regress out the 65 proxies for hidden
confounding and $X',Y'$ are the corresponding residuals. The response
variable is the expression of one (randomly chosen) gene while the covariates
comprise all other expressions.
If there is hidden (linear) confounding and the proxy variables indeed
capture the true underlying hidden 
confounding variables, that is, the linear span of the 65 proxy variables
equals the linear span of the unobserved hidden variables, the $X',Y'$ data
are unconfounded. This in turn would imply that the deconfounded Lasso as
in \eqref{TrimLasso} would
give similar results on $(X,Y)$ and $(X',Y')$, while this would not be the
case for the plain Lasso as it would be subject to some bias when running
it on the confounded data $(X,Y)$. We aim here to illustrate
that the deconfounded Lasso is indeed more robust than the plain Lasso when
estimation 
is done once on the original $(X,Y)$ and once on the approximately deconfounded
data $(X',Y')$.
\begin{figure}[!htb]
  \begin{center}
    \includegraphics[scale=0.13]{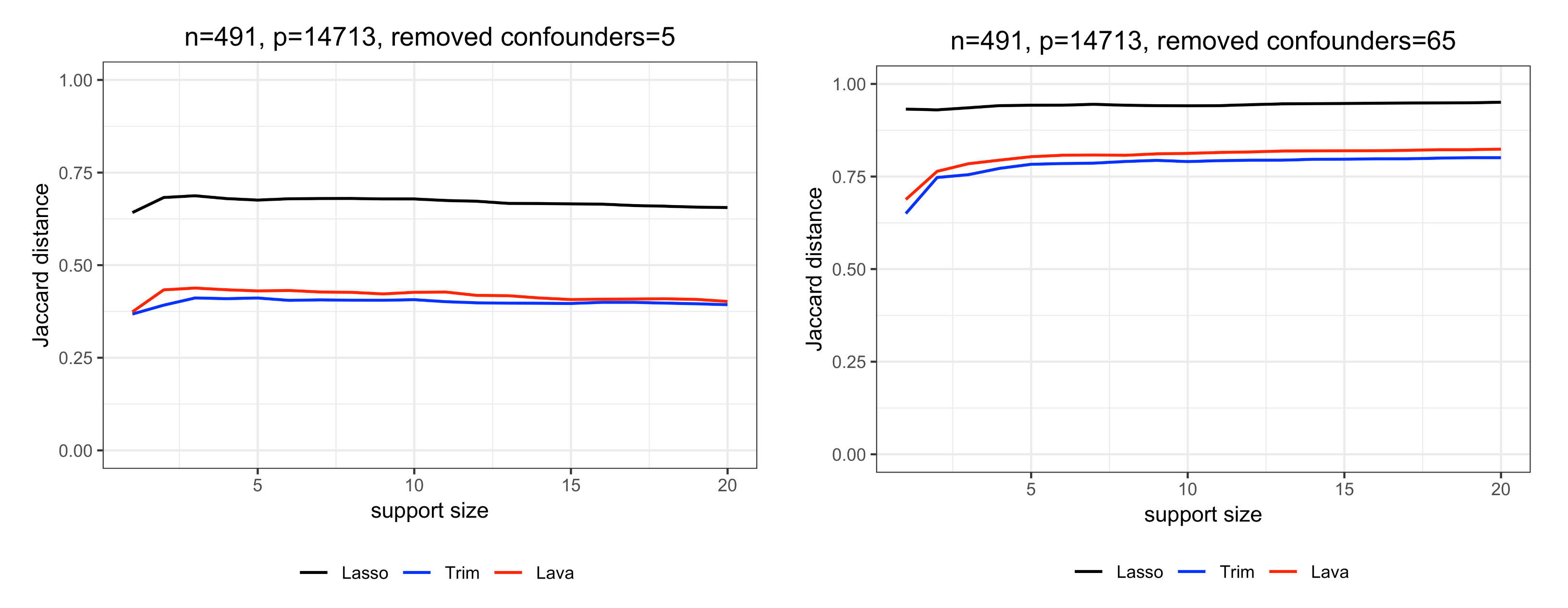}
    \caption{Stability of selected features in GTEx data. Jaccard distances
      for sets of selected variables on original 
      and proxy adjusted (for approximate deconfounding) data. Adjustment
      with 5 (left) and 65 (right) proxies for hidden confounding. x-axis:
      number of top selected features (support size); y-axis: Jaccard distance
      between sets of top selected features (of fixed support sizes) based
      on original and proxy adjusted data. Black: plain Lasso; blue:
      trim transform followed by Lasso; red: Lava. The figure is essentially taken from \citet{cevid2018spectral}.}\label{fig.jaccard}
  \end{center}
  \end{figure}
  
The response variable is one randomly selected gene and the remaining $p = 14'712$ gene expressions are the covariates. We compute the regularization paths of
the  Lasso with the Trim transformed data and of the plain Lasso: this
leads to sets of active variables with non-zero coefficient estimates
$\hat{S}_{\mathrm{TrimLasso}}^{(r)}(\lambda) \subset \{1,\ldots ,p\}$ and
  $\hat{S}_{\mathrm{Lasso}}^{(r)}(\lambda) \subset \{1,\ldots ,p\}$, where
  $r = 1,2$ denotes the original $(X,Y)$ and the proxy adjusted data
  $(X',Y')$, respectively. As a measure of robustness or consensus, we
  compute the 
  Jaccard distance between $r = 1$ and $r=2$ for sets of the same
  cardinality (by varying $\lambda$). For comparison, we consider also the Lava
  estimator with the tuning parameter as in \eqref{lambda2lava}. Figure
  \ref{fig.jaccard} reports the results when  
  adjusting once with $5$ and once with all $65$ proxy confounding
  variables, and averaging 
  over 500 randomly chosen response variables. The problem is very
  high-dimensional and with a high noise level: nevertheless, we clearly
  see that deconfounding with the Trim transform, and Lava
  likewise, provide more robustness than the plain Lasso, across a
  large range of cardinalities of the active sets. We note that the
  robustness or consensus decreases with more proxy adjustment: this is
  mainly due to 
  the fact that the data sets $(X,Y)$ and $(X',Y')$ become more different
  with more adjustment. But the advantage remains when considering differences
  between the methods.

  We mention here that \citet{shah2018rsvp} provide vaguely related results on
  robustness for the GTEx data for another Ridge-type procedure for
  undirected graphical models.

 \subsection{The doubly debiased Lasso}

 Assigning uncertainty is a core task in statistical inference. Substantial
 progress has been made for 
 low-dimensional parameters in high-dimensional models. The prime example
 is about inference for single components of a high-dimensional regression
 parameter. The 
 debiased or desparsified Lasso has become a basic machinery for
 constructing hypothesis tests and confidence intervals
 \citep{zhangzhang11,vdgetal13}, see also \citet{dezetal15} for some
 review of the earlier work. An interesting 
 property of the debiased or desparsified Lasso is its efficiency,
 assuming sparsity conditions \citep{vdgetal13}.

 We briefly review here the approach of \citet{guoetal2020} on the doubly
 debiased Lasso to obtain hypothesis tests and confidence intervals for
 single regression coefficients $\beta_j^0$ in model \eqref{SEMlin} in
 presence of hidden confounding.

\paragraph{The standard debiased Lasso.}
 The idea of debiasing the Lasso is based
 on partial regression. For ordinary least squares estimation in the $p <
 n$ regime, we obtain the
 estimator $\hat{\beta}_{\mathrm{OLS};j}$ as follows:
 \begin{eqnarray}\label{partialregr}
   & &\mbox{do least squares regression of $X^{(j)}$ versus $X^{(-j)}$ and
       denote the residuals by $Z^{(j)}$},\nonumber\\
   & &\hat{\beta}_{\mathrm{OLS};j} = (Z^{(j)})^T Y/\|Z^{(j)}\|_2^2 = (Z^{(j)})^T
       Y/((Z^{(j)})^T X^{(j)}),
 \end{eqnarray}
 where $X^{(-j)}$ is the $(n \times (p-1))$ matrix arising from deleting the
 $j$th column of $X$. 
 The first regression in \eqref{partialregr} is ill-posed if $p > n$. 
 The debiased Lasso then uses instead
 \begin{eqnarray}\label{lassoX}
   \mbox{a Lasso regression of $X^{(j)}$ versus
   $X^{(-j)}$ and denote the residuals again by $Z^{(j)}$}.
 \end{eqnarray}
When using them in the second regression we obtain
\begin{eqnarray*}
 \frac{(Z^{(j)})^T Y}{(Z^{(j)})^T X^{(j)}} = \beta_j + \sum_{k \neq j}
  \frac{(Z^{(j)})^T X^{(k)} \beta_k}{(Z^{(j)})^T X^{(j)}} + \frac{(Z^{(j)})^T
  \eps_Y}{(Z^{(j)})^T X^{(j)}},
\end{eqnarray*}
where we assume an unconfounded model $Y = X \beta + \eps_Y$ with
$\EE[\eps_Y|X] = 0$.
Unlike as for least squares, $(Z^{(j)})^T X^{(k)} \neq 0$ for $k \neq j$
and hence there is a bias term
\begin{eqnarray*}
  B = \sum_{k \neq j} \frac{(Z^{(j)})^T X^{(k)} \beta_k}{(Z^{(j)})^T
  X^{(j)}}.
\end{eqnarray*}
An obvious estimator for the bias arises by plugging in a Lasso estimate of $Y$
versus $X$, resulting in
\begin{eqnarray}\label{hatbias}
  \hat{B} = \sum_{k \neq j} \frac{(Z^{(j)})^T X^{(k)}
  \hat{\beta}_{\mathrm{Lasso};k}}{(Z^{(j)})^T 
  X^{(j)}},
\end{eqnarray}
and the debiased or desparsified Lasso is then defined as
\begin{eqnarray*}
  \hat{\beta}_{\mathrm{debiasedLasso};j} = \frac{(Z^{(j)})^T Y}{(Z^{(j)})^T
  X^{(j)}} - \hat{B}.
\end{eqnarray*}

In case of hidden confounding in model \eqref{SEMlin}, both regressions in
\eqref{partialregr} and \eqref{lassoX} are exposed to bias from hidden
confounding and standard methodology does not work. Following the ideas
developed in Sections 
\ref{subsec.spectrans}-\ref{subsec.Trimlasso}, we propose to Trim
transform the data twice, once before applying the Lasso in the $X^{(j)}$
versus $X^{(-j)}$ regression
\eqref{lassoX} and once before using the Lasso in $Y$ versus $X$ for being
plugged-in to the 
bias estimator in \eqref{hatbias}. By doing so, we remove bias thanks to
spectral transformations and hence the words ``doubly debiased''. 
Of course, there are tuning parameters
to be chosen, namely for each of the Lasso regressions appearing in
\eqref{lassoX} and \eqref{hatbias}. This issue is analogous as
for the standard debiased or desparsified Lasso but perhaps one tick more
difficult as indicated in Section \ref{subsec.regularization}. The details
are given in \citet{guoetal2020} and the resulting estimator is called the
doubly debiased Lasso $\hat{\beta}_{\mathrm{DDLasso}}$. 

The following result holds.
\begin{fact} \citep{guoetal2020}
Consider the confounding model in \eqref{SEMlin} with $\max_j \Sigma_{jj} =
\mathcal{O}(1)$, where $\Sigma = \Cov(X)$.  Under similar conditions as in (A1)-(A3)
and assuming sparsity for both the regressions of $Y$ versus $X$ and the
one in \eqref{lassoX},  
\begin{eqnarray*}
& &V_j^{-1/2} (\hat{\beta}_{\mathrm{DDLasso};j} - \beta^0_j)
    \Longrightarrow {\cal N}(0,1)\ \ (p \ge n \to \infty),\\
  & &V_j = n^{-1} \Var(\eps_Y)g_j(X),
\end{eqnarray*}
with a known specific function $g_j(X)$ of the design matrix $X$ which
is of order 1 as $p \ge n \to \infty$.
In addition, if the trimming threshold is such that the fraction of the
shrunken singular values converges to zero (only the 
very large singular values are Trimmed to the corresponding quantile
value), the doubly debiased Lasso is as efficient as the ordinary least
squares estimator in low dimensional settings, that is 
\begin{eqnarray*}
V_j \asymp n^{-1} \Var(\eps_Y) (\Cov(X))^{-1}_{jj}. 
\end{eqnarray*}
  \end{fact}

\subsubsection{An illustration on the GTEx data}\label{subsec.GTEx2}

We consider again the GTEx data mentioned in Section \ref{subsec.GTEx1},
but now with a somewhat smaller dimensionality $p = 12'646$ but an
increased sample size of 
$n = 706$ (removing some of the covariates with missing values due to
larger sample size). There are approximately one thousand landmark
genes of particular importance and interest.

\begin{figure}[!htb]
  \begin{center}
    \includegraphics[scale=0.7]{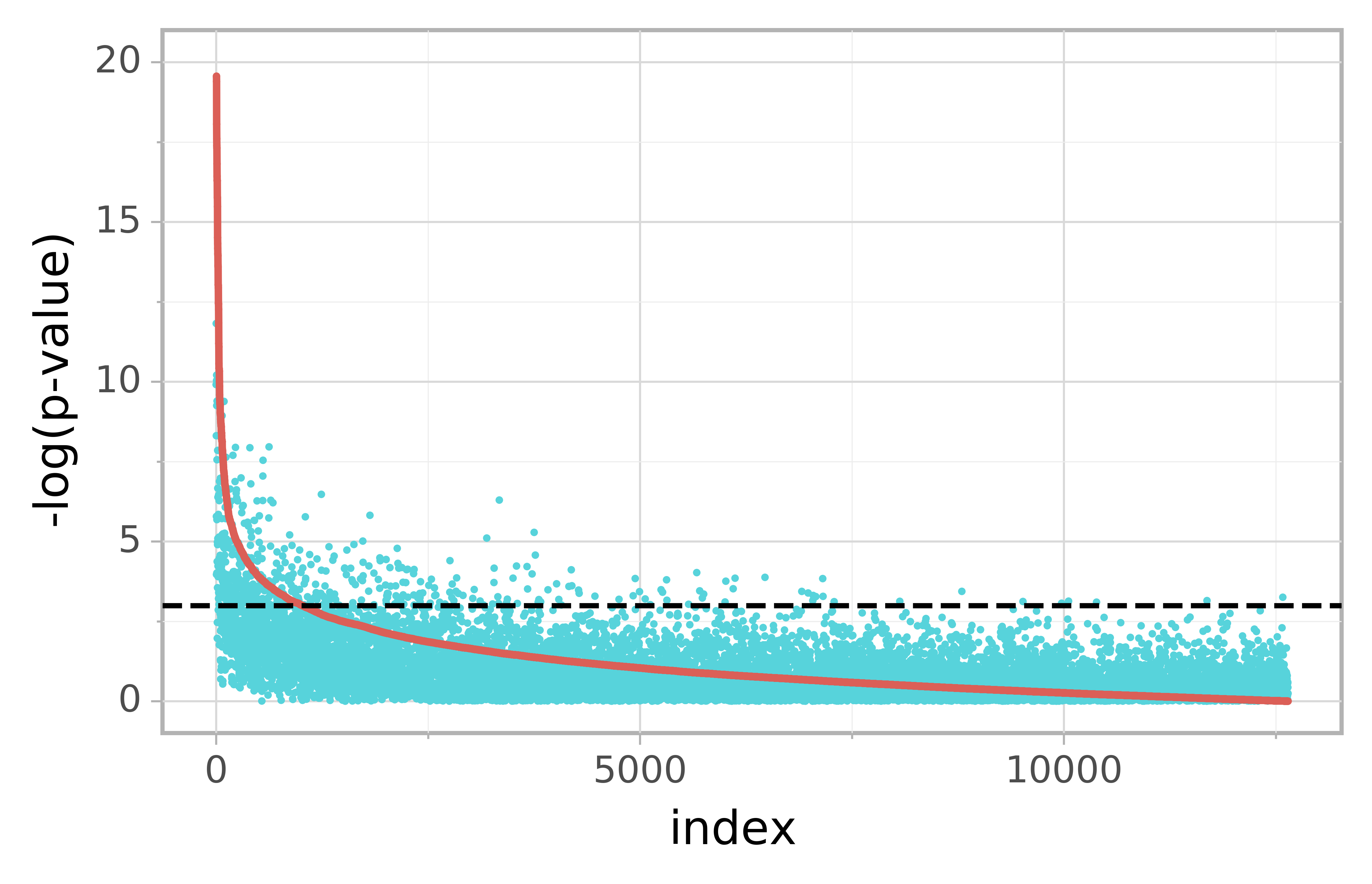}
    \caption{$P$-values for two-sided test of the hypothesis $H_{0,j}:
      \beta_j^0 = 0$ with two-sided alternative for GTEx data. Doubly
      debiased Lasso 
      (blue) and debiased Lasso (red) for the expression of a predetermined
      landmark gene (being the response $Y$) 
versus all other gene expressions (being the covariates $X$). x-axis: index
of covariates, ordered by decreasing significance based on the
Debiased Lasso.; y-axis: negative log $P$-value. Black dotted line indicates
the 5\% significance level with the value $-
\log(0.05)$. The figure is essentially taken from
\citet{guoetal2020}.}\label{fig.orderingpval}   
  \end{center}
  \end{figure}

Figure \ref{fig.orderingpval} illustrates the difference between the $P$-values from the Doubly
Debiased Lasso in comparison to the plain debiased Lasso which does not
protect against potential hidden confounding. The plot considers a
particular landmark gene whose expression is the response variable and all
other 12'645 gene expressions are covariates.
The Doubly debiased Lasso
claims less significance which seems a plausible finding (and it is not
primarily due to larger variance which is not shown here).
  
\begin{figure}[!htb]
  \begin{center}
    \includegraphics[scale=0.5]{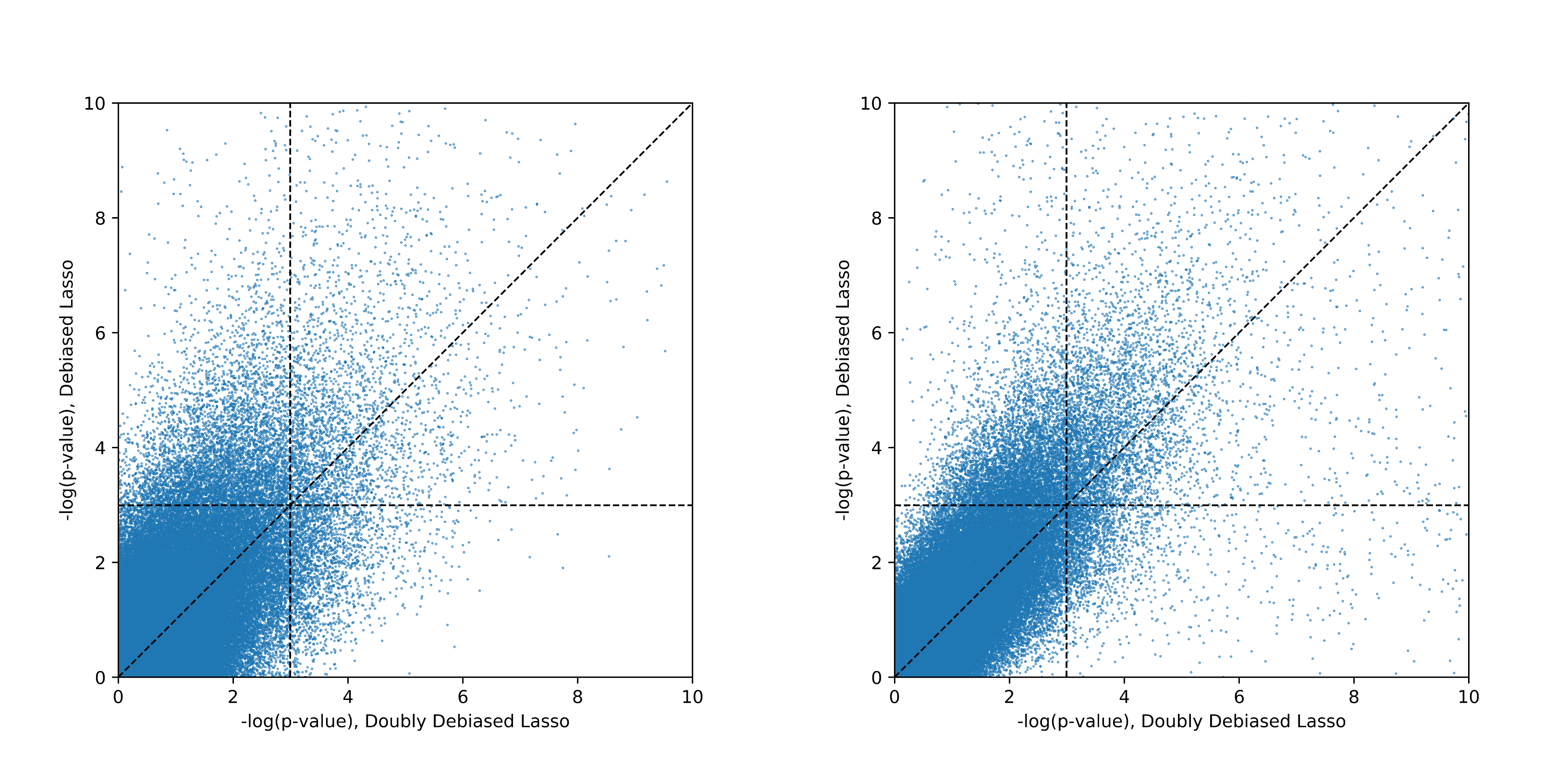}
    \caption{Stability of $P$-values in GTEx data. Comparing $P$-values from
      original data and its proxy 
      adjusted version for approximate deconfounding. Two-sided tests
      of the hypothesis $H_{0,j}:\ \beta_j^0 = 0$, for 10 landmark gene
      expression as responses and all other expressions as covariates,
      obtained by doubly debiased Lasso (x-axis, - log $P$-value) and
      standard debiased Lasso (y-axis, - log $P$-value). Original $(X,Y)$
      data (left) 
      and adjusted $(X',Y')$ with 65 proxies for hidden confounding
      (right). Horizontal and vertical black dashed lines indicate the 5\%
      significance level. The figure is taken from \citet{guoetal2020}.}\label{fig.scatterpval}  
  \end{center}
  \end{figure}

We also illustrate increased robustness of the doubly debiased Lasso. As
explained already in Section \ref{subsec.GTEx1}, there are 65 additional
proxy variables 
which are aimed to approximate unobserved hidden confounding. Figure \ref{fig.scatterpval}
shows $P$-values for 10 response landmark genes (and the plots comprise all $P$-values from the 10 regressions). We can see from the left plot
that the doubly debiased Lasso is much more conservative for the
potentially confounded original $(X,Y)$ data.  The cloud of points is skewed
upwards showing that the
standard debiased Lasso declares many more predictors as significant. On
the other hand, in the right plot the $P$-values obtained by the two methods
are much more similar for the proxy-adjusted unconfounded $(X',Y')$ data and the
point cloud is now  
much less skewed upwards. The remaining deviation from the line $y=x$ might be
due to the remaining confounding, not accounted for by regressing out the given
confounder proxies.  
Figure \ref{fig.scatterpval} describes the results.

\subsection{When dense confounding fails}\label{subsec.fails}

When hidden confounding is substantial but fails to be dense in the sense
that it 
affects many of the components of $X$, the deconfounding Trim
transformation technique does not 
effectively remove the entire bias and the parameter $\beta^0$ is not
identifiable from the data generating distribution. In such situations,
other assumptions are required, see also Section \ref{sec.anchor}. 

However, deconfounding leads to robustification, as mentioned in Section
\ref{subsec.robustify}: the points (1)-(4) there are relevant in general,
also for inference with the doubly debiased Lasso. We repeat here again that
deconfounding can lead to substantial improvements and seems
to never make things substantially worse than not taking any action against
hidden confounding. The price to be paid for deconfounding is typically a
slightly larger variance of the estimator resulting in a somewhat reduced 
efficiency for data without any confounding.

\section{Anchor regression: towards causality, distributional robustness and distributional replicability}\label{sec.anchor}  

We consider now the general situation from Figure \ref{fig1}: it extends Figure
\ref{fig2} since the directions between the variables $X, Y, H$ are
unknown. Furthermore, we abandon here the major assumption of dense
confounding from Section \ref{sec.deconfounding}. Both are
important relaxations in 
practice. However, this comes with the price of requiring access to
exogenous variables $A$ as indicated in Figure \ref{fig1}: as an
example, we mention the case where the variables $A$ represent mean shift
perturbations (see Figure \ref{fig1}), where exogeneity (source node in the
graph in Figure \ref{fig1}) is often a reasonable
assumption. 

Instrumental variables regression is a popular proposal for a special case
with perturbations
\citep{bowdenturk90,angristetal96,stock2003,imbens2014,imru15}. The SEM
in \eqref{SEMlin} is extended to
\begin{eqnarray}\label{SEMIV}
  & &Y_i \leftarrow X_i \beta^0 + H_i \delta + \eps_{Y,i},\nonumber\\
  & &X_i \leftarrow A_i \kappa + H_i \gamma + \eps_{X,i},\nonumber\\
  & &A_i,\ H_i,\ \eps_{Y,i},\ \eps_{X,i}\ \mbox{jointly independent},
\end{eqnarray}
where the random variables are i.i.d. across $i=1,\ldots ,n$. The main
assumption here is that the so-called instrumental variables $A_i$ do not
directly affect the hidden variable $H_i$ nor the response variable $Y_i$. The
well-known two-stage least squares estimator is then defined as the least
squares estimator on linearly transformed data:
\begin{eqnarray}\label{2SL}
  & &\tilde{Y} = \Pi_A Y,\ \tilde{X} = \Pi_A X,\ \Pi_A = A (A^TA)^{-1}
      A^T,\nonumber\\
  & &\hat{\beta}_{\mathrm{TSLS}} = \argmin_{\beta} \|\tilde{Y} - \tilde{X}
      \beta\|_2^2/n = \argmin_{\beta} \|\Pi_A(Y - X \beta)\|_2^2/n. 
\end{eqnarray}

\subsection{Anchor regression}
More generally than two-stage least squares estimation in \eqref{2SL}, we
can look at its regularized version, called anchor regression
\citep{rothenhetal18}:
\begin{eqnarray}\label{anchorest} 
\hat{\beta}_{\mathrm{anchor}}^{(\gamma)} = \argmin_{\beta} \|(I - \Pi_A)(Y - X
  \beta)\|_2^2/n + \gamma \|\Pi_A(Y - X \beta)\|_2^2/n,
\end{eqnarray}
for some regularization parameter $0 \le \gamma \le \infty$. With $\gamma =
0$, we obtain adjustment with respect to $A$, i.e., partialling out the
linear effect of $A$, $\gamma = 1$ corresponds to ordinary least squares
and $\gamma = \infty$ is two-stage least squares. This regularization has been
proposed much earlier in a different but equivalent form under the name of
K-class estimators, mainly for reducing the large (or infinite) variance of
two-stage 
least squares for estimation of $\beta$ \citep{theil58,jakjon20}. The
computation of anchor 
regression is extremely easy and modular: one can simply transform the
data
\begin{eqnarray*}
  & &\tilde{X} = W_{\gamma}X,\ \tilde{Y} = W_{\gamma} Y,\\
  & &W_{\gamma} = I - (1 - \sqrt{\gamma}) \Pi_{A},
  \end{eqnarray*}
and then use least squares estimation of $\tilde{Y}$ versus
$\tilde{X}$. One can also 
consider sparsity-regularized anchor regression with e.g. the
$\ell_1$-norm penalty:
\begin{eqnarray*}
\hat{\beta}_{\mathrm{anchor}}^{(\gamma)} = \argmin_{\beta} \|(I - \Pi_A)(Y - X
  \beta)\|_2^2/n + \gamma \|\Pi_A(Y - X \beta)\|_2^2/n + \lambda
  \|\beta\|_1
\end{eqnarray*}
which can be solved by running standard Lasso of $\tilde{Y}$ versus
$\tilde{X}$.

The anchor regression method is also called causal regularization since it
regularizes least squares towards the causal parameter (whereas the
motivation for \eqref{anchorest} above has been to regularize the TSLS
estimator towards least 
squares to reduce variance). For $\gamma \to \infty$ we approximate a
causal solution under the assumptions of instrumental variables
regression. More generally, one can improve robustness and replicability
when choosing $\gamma$ clearly larger than $1$ as  discussed in the
sequel.

We can also connect anchor regression to deconfounding from Section
\ref{sec.deconfounding}. When taking the anchor variables $A$ as the first
$\hat{q}$ principal components of $X$, then $\gamma = 0$
corresponds to PCA adjustment as described in \eqref{pcashrink}; for
$\gamma >0$ but small, such an anchor regression would shrink the first
$\hat{q}$ singular values of $X$ but it is not a spectral transform any
longer of the form as in \eqref{pcaspec} with any transformed singular
values $\tilde{d}_i$.
Also, when using $A$ as the first principal components 
of $X$, the exogeneity assumption as in Figure \ref{fig1} is violated, a
crucial condition for what we discuss next.

\subsection{Distributional robustness of Anchor regression}\label{subsec.distrranchor}
\citet{rothenhetal18} take a very different view of \eqref{anchorest} than
improving the mean squared error of the two-stage least squares estimator
\eqref{2SL} in IV regression, namely that
anchor regression is sensible even when the main assumptions of IV
regression fail. That is, if we allow that $A$ directly affects the hidden
variables $H$ or the response $Y$
in \eqref{SEMIV}, which implies that $\beta^0$ is not identifiable from the
data, anchor regression is estimating an interesting parameter, as we
discuss next. Consider the population version of anchor regression in
\eqref{anchorest}:
\begin{eqnarray}\label{anchorpop}
\beta^{(\gamma)} = \argmin_{\beta} \EE[|(I - P_A)(Y_i - X_i \beta)|^2] +
  \gamma \EE[|P_A(Y_i - X_i \beta)|^2],
\end{eqnarray}
where $P_A(\cdot) = \EE[\cdot|A]$ is the population version of $\Pi_A$ under
a linearity assumption as in \eqref{mod.anchor} below, and the index $i$ is
arbitrary (since we 
assume that the data is i.i.d. across samples). This population parameter
$\beta^{(\gamma)}$ is a \emph{regularized} population parameter, where the
regularization is not used to obtain better statistical finite sample
properties.
Instead, the regularization has a direct relation to distributional robustness.

To explain such a robustness, assume that the training data are
i.i.d realizations of the following structural equation model:
\begin{eqnarray}\label{mod.anchor}
\begin{pmatrix}
           X_i \\
           Y_i \\
           H_i
         \end{pmatrix}^T = B \begin{pmatrix}
           X_i \\
           Y_i \\
           H_i
         \end{pmatrix}^T + \eps_i + M A^T_i = (I - B)^{-1} (\eps_i + M A^T_i),
\end{eqnarray}
where (all the components of) $A_i, \eps_i$ are jointly independent and $M$
is a coefficient matrix of dimension $(\mbox{dim}(X_i,Y_i,H_i) \times
\mbox{dim}(A_i))$. 
Note that
$I - B$ is always invertible if the model structure corresponds to an
acyclic directed graph.

We define the system under shift perturbations $v$ by the same equations as
in \eqref{mod.anchor} but replacing the term $M A$ from the contributions of
the anchor variables by a deterministic or stochastic perturbation vector
$v$. That is, the system under shift perturbations satisfies: 
\begin{eqnarray}\label{mod.pertanchor}
\begin{pmatrix}
           X^v \\
           Y^v \\
           H^v
         \end{pmatrix}^T = B \begin{pmatrix}
           X^v \\
           Y^v \\
           H^v
         \end{pmatrix}^T + \eps + v = (I-B)^{-1} (\eps + v), 
\end{eqnarray}
with $\eps$ having the same distribution as $\eps_i$ in
\eqref{mod.anchor}. 
The shift vector $v$ is assumed to be in the span of $M$, that is $v = M
\delta$ for some vector $\delta$. Thus, the vector $v$ shifts the variables
$X^v, Y^v, H^v$ in the same direction as $A_i$, according to the range (or
span) of $M$ but with possibly different strengths. The
variables $X^v, Y^v$ can be 
interpreted as the test data coming from a different distribution than the
training data from model \eqref{mod.anchor}.

\medskip\noindent
\emph{An example with discrete anchors, encoding different environments.} \\
We often have the following situation in mind. The data are heterogeneous
from various subpopulations or environments labeled by $\{1,\ldots
,\ell\}$. 
These are then encoded with
$\ell$-dimensional anchor variables $A_i$ in the form of dummy variables. The
heterogeneity of the data enters as distributional additive 
shifts (or perturbations) in terms of $M A_i^T$, As an environment, it is
often reasonable to assume that $A_i$ is exogenous, i.e., a source node in
the graph in Figure \ref{fig1}. The data generated by
\eqref{mod.pertanchor} is typically the test data where (realizations) of
the anchor variable $A$ is not available. 

\medskip
The 
following result holds:
\begin{fact}\label{fact3}\citep{rothenhetal18}
  Consider random variables $X_i,Y_i$ as in \eqref{mod.anchor} and
  $X^v,Y^v$ as in \eqref{mod.pertanchor}. Then, for any $b \in \R^p$ it
  holds that 
  \begin{eqnarray*}
    \ \ \sup_{v \in C_{\gamma}} \EE[(Y^v - X^v b)^2] = \EE[((I - P_A)(Y_i -
    X_i^T b))^2] + \gamma \EE[(P_A(Y_i - X_i^T b))^2],
  \end{eqnarray*}
  where
\begin{eqnarray}\label{shift-pertclass}
  C_{\gamma} = \{v;& &\hspace*{-4mm}v = M \delta\ \mbox{for random or
                       deterministic}\ 
  \delta,\ \mbox{uncorrelated with}\ \eps\nonumber\\
  & &\hspace*{-4mm}\mbox{and}\ \EE[\delta \delta^T]
  \preceq \gamma \EE[A A^T]\}.
\end{eqnarray}
\end{fact}
Fact \ref{fact3} establishes distributional robustness of the population
version of anchor regression: the parameter $\gamma$ has an exact
correspondence to the class $C_{\gamma}$ of shift perturbations. From a
practical view-point, Fact \ref{fact3} tells us that we can construct an
estimator on the training data only, by employing causal regularization,
which protects on new test data which arises from shift perturbations as in
\eqref{shift-pertclass}.

\citet{rothenhetal18} give finite sample versions of the result in Fact
\ref{fact3} and show empirical examples how prediction can be improved
thanks to distributional robustness: if the test data is a perturbed
version of the training data, formalized with $X_i^v,Y_i^v$ for some
perturbation vector $v$, then the expected worst case squared error loss on
the test data can be optimized by anchor regression.

\paragraph{Choosing $\gamma$ and specifying anchor variables.}
The choice of $\gamma$ in anchor regression or causal regularization can be
addressed from different angles. If we want to insure ourselves
against bad perturbations in span($M$) of a certain size, as defined in
\eqref{shift-pertclass} and aiming for worst case optimal prediction, then
$\gamma$ corresponds to the multiplication factor of the observed
heterogeneity in the data. That is, e.g. $\gamma = 5$
corresponds to perturbations $\sqrt{5}$ times as large as the ones we have
observed in the data. Alternatively, we can consider leave-one-environment-out
cross-validation and choose $\gamma$ which optimizes the worst case
performance among the left-out environments (being the test data).  

Regarding the specification of anchor variables, as mentioned above, $A_i$ 
should be exogenous: we describe below an example with heterogeneity
arising from different environments. The general idea is to
\emph{stabilize} the estimator 
$\hat{\beta}_{\mathrm{anchor}}$ over the values of $A_i$ by taking a
large value of $\gamma$ which enforces that the residuals are nearly
orthogonal to $A_i$; see also \citet{pfisteretal19} and a replicability
result in Fact \ref{fact4}. This is the opposite action
than using $A_i$ as an additional covariate which would correspond to
$\gamma = 0$. Exogeneity and $\gamma = \infty$ (two stage least squares
estimation) plays also a prominent role in IV regression for deconfounding the
effects of hidden confounders $H_i$. Thus, the stabilizing anchor
regression estimator is ideally pursued with exogenous anchor variables
and large values of $\gamma$.


\subsection{Distributional replicability and external validity with anchor regression}\label{subsec.dilutedcausal}

Anchor regression also leads to an improved replicability on new data, say
from a related study. We argue here that the parameter
\begin{eqnarray*}
  \beta^{(\to \infty)} = \limsup_{\gamma \to \infty} \beta^{(\gamma)}
\end{eqnarray*}
can be replicated on new data from a different distribution than the
training data. In view of Fact \ref{fact3}, $\beta^{(\to \infty)}$ leads to
distributional robustness for arbitrarily large perturbations $v = M
\delta$ in the span of $M$. If the assumptions from instrumental variables
regression hold, then $\beta^{(\to \infty)} = \beta^0$ which is the causal
parameter, but that's not the case in general. The causal
parameter has an invariance property with respect to certain arbitrarily
strong perturbations. Also
$\beta^{(\to \infty)}$ leads to an invariance of the residuals, namely: 
\begin{eqnarray}\label{invres}
  & &Y^v - X^v \beta^{(\to \infty)}\nonumber\\
  & &\mbox{has the same distribution for all arbitrarily strong
      perturbations $v$ as in \eqref{mod.pertanchor}}.
  \end{eqnarray}
According to a general relation between causality and
invariance, and due to the residual invariance from \eqref{invres}, we call
$\beta^{(\to \infty)}$ the ``diluted causal'' parameter.    
 
We consider now the following setting. The first dataset is generated from the
model \eqref{mod.anchor} whose distribution induces the diluted causal
parameter $\beta^{(\to 
  \infty)}$ (being a function of the data generating probability
distribution). The second (test or validation) dataset is generated from a
perturbed version as in model \eqref{mod.pertanchor} whose distribution
generates the diluted causal parameter $b^{'(\to \infty)}$.

For our replicability result in Fact \ref{fact4}, we require the so-called projectability assumption:
 \begin{eqnarray}\label{projectab}
    I = \{\beta; \EE[Y - X\beta|A] \equiv C\} \neq \emptyset\ \mbox{for any
   constant C}.
  \end{eqnarray}
  This condition holds if and only if
  \begin{eqnarray*}
    \mbox{rank}(\mathrm{Cov}(A,X)) = \mbox{rank}[\mathrm{Cov}(A,X),
    \mathrm{Cov}(A,Y)],
  \end{eqnarray*}
where $[\mathrm{Cov}(A,X), \mathrm{Cov}(A,Y)]$ denotes the extended matrix by
concatenating the columns of the two matrices. For example, if
$\mbox{rank}(\mathrm{Cov}(A,X))$ is full rank and $\mathrm{dim}(A) \le 
\mathrm{dim}(X)$, the projectability condition \eqref{projectab} holds.
\begin{fact}\label{fact4} \citep{rothenhetal18}
Consider the diluted causal parameters $\beta^{(\to \infty)}$ from model 
\eqref{mod.anchor} and $b^{'(\to \infty)}$ from
model \eqref{mod.pertanchor}. Assume the projectability condition
\eqref{projectab} for the model \eqref{mod.anchor}. Then, $\beta^{(\to \infty)}
= b^{'(\to \infty)}$, that is, the diluted causal parameter is replicable
on the new perturbed dataset. 
\end{fact}

\subsubsection{An illustration on the GTEx data}
We illustrate the distributional replicability for the diluted causal
parameter $\beta^{(\to \infty)}$ on the GTEx data mentioned already in Section
\ref{subsec.GTEx1} and \ref{subsec.GTEx2}. 

Here, we consider 13 different tissues for which $p = 12'948$ gene expression
measurements and $65$ proxies of confounding are measured. The 13 different
tissues correspond to 13 different datasets consisting of response
variables $Y$ being one of the gene expressions, covariates $X$ comprising
all other gene expressions and anchor variables $A$ being the 65 proxy
variables. The sample size varies between $300-700$ across the 13
tissues.
\begin{figure}[!htb]
  \begin{center}
\includegraphics[scale=0.5]{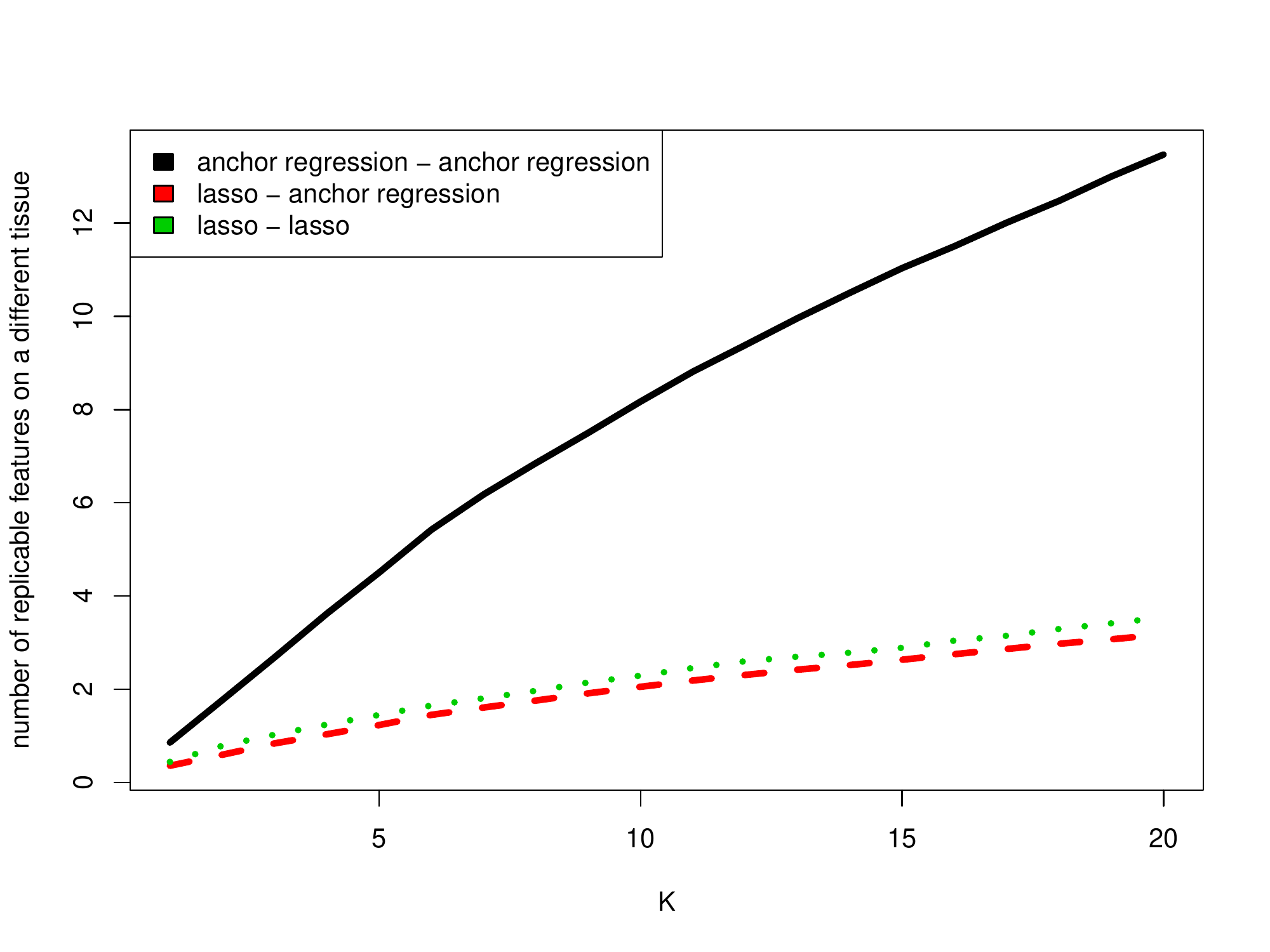}
\caption{Replicability of diluted causal parameter $\beta^{(\to \infty)}$ on
  GTEx data. x-axis: number $K$ of top ranked features; y-axis: average
  overlap (number) among 
  the top K features among $13 \choose 2$ tissue pairs, and averaged over
  200 randomly selected response variables each being one of the available gene
expressions. Anchor regression with $\gamma = 16$ on both tissues in each
pair (solid black), Lasso on both tissues in each pair (dotted green), anchor
regression 
with $\gamma = 16$ on one tissue and Lasso on the other tissue for each
tissue pair (dashed red). The figure is taken from
\citet[Fig.4]{rothenhetal18}.}\label{fig.replicabanchor}
\end{center}
\end{figure}

We consider anchor regression with $\gamma = 16$ (being chosen as a large
value, yet still improving the variance in comparison to choosing $\gamma =
\infty$, i.e., two-stage least squares) and cross-validated choice
of the tuning parameter $\lambda$ for an $\ell_1$-norm penalty as an
estimator for $\beta^{(\to \infty)}$, for each of the 13 tissues
(datasets). The goal is to evaluate the degree of replicability and
external validity of the anchor regression estimator. Figure
\ref{fig.replicabanchor} illustrates the results. The anchor regression
estimate $\hat{b}^{t,(\gamma = 16)}$ for one tissue $t$ is compared with
another one $\hat{b}^{t',(\gamma = 16)}$ for another tissue $t'$. The
overlap (number) among the top $K$ variables (features), according to the
absolute value of the estimates, is counted and averaged over all $13
\choose 2$ pairs of tissues (datasets) and 200 random choices
of a response as one of the available gene expressions. Figure
\ref{fig.replicabanchor} displays the results. There is some
evidence for the GTEx data that indeed, anchor regression for the diluted
causal parameter $\beta^{(\to \infty)}$ has higher degree of replicability on
new perturbed datasets. 

The interpretation of the diluted causal parameter $\beta^{(\to \infty)}$ is,
however, different from the usual least squares parameter, and leads to
invariance of residuals as described in \eqref{invres}. 

%

\section{Discussion}

\paragraph{Extensions.}
We have explained here the concepts for linear models only. Modifications for
generalized linear models or nonlinear models are certainly of interest. In
the context of nonlinear anchor regression, some methodological and
algorithmic proposals have been illustrated empirically in \citet{pb20}.
In general, for models with nonlinear regression functions, we can view the
proposed methods as to perform deconfounding or distributionally
robustifying the linear component of a general regression function. In fact,
from a transfer learning perspective, for replicability on 
new data, it seems hard to go beyond linear extrapolation for strong
perturbations arising in new data \citep{christetal20}.

Robustification and stabilizing over different environmental
conditions or different datasets from a ``causal structural equation
model'' point of view has been worked out also for independent component
analysis \citep{pfisterweich2019} or dynamical systems modeling \citep{pfisteretal2019}.   

\paragraph{Summary.}
We have argued that deconfounding or causal regularization (i.e., anchor
regression) are powerful tools for improving replicability or
distributional robustness; see Sections \ref{subsec.robustify},
\ref{subsec.GTEx2}, \ref{subsec.distrranchor} and
\ref{subsec.dilutedcausal}. For linear systems, the
operational procedures are extremely simple and modular: it's just linearly
pre-transforming the data and then using any modern regression technique on
such transformed data. Such pre-transformations are also crucial for the
important issue of ``attribution'' \citep{efron2019,efron2020}: for
high-dimensional densely confounded linear models, the doubly debiased Lasso
\citep{guoetal2020} leads to hypothesis tests and confidence intervals for
the unconfounded (causal) parameter and thus, as an important consequence, to
improved replicability even though the data is corrupted by latent
perturbations or ``(context) drifts'' \citep{efron2020}.

%
%
%

\paragraph{Acknowledgments.} The research of P. B\"{u}hlmann and D. \'Cevid
was supported by the European Research Council under the Grant Agreement No 786461 (CausalStats - ERC-2017-ADG).

\bibliographystyle{apalike}
\bibliography{references}

\end{document}